# Electrical characterization of AMS aH18 HV-CMOS after neutrons and protons irradiation

**D M S Sultan[a,1], Sergio Gonzalez Sevilla[a], Didier Ferrere[a], Giuseppe Iacobucci[a], Ettore Zaffaroni[a], Winnie Wong[a], Mateus Vicente Barrero Pinto[a], Moritz Kiehn[a], Mridula Prathapan[b], Felix Ehrler[b], Ivan Peric[b], Antonio Miucci[c], John Kenneth Anders[c], Armin Fehr[c], Michele Weber[c], Andre Schoening[d], Adrian Herkert[d], Heiko Augustin[d], and Mathieu Benoit[a]**

[a] *Department of Nuclear Physics, University of Geneva,*
*24, quai Ernest-Ansermet, CH-1211 Genève 4, Switzerland*

[b] *Institute of Data Processing and Electronics, Karlsruhe Institute of Technology,*
*P.O. Box 3640, 76021 Karlsruhe, Germany*

[c] *Physics Institute, University of Bern,*
*Sidlerstrasse 5, 3012 Bern, Switzerland*

[d] *Im Heuenheimer Feld, Physics Institute, University of Heidelberg,*
*69120 Heidelberg, Germany*

*E-mail*: dms.sultan@unige.ch

ABSTRACT: In view of the tracking detector application to the ATLAS High Luminosity LHC (HL-LHC) upgrade, we have developed a new generation of High Voltage CMOS (HV-CMOS) monolithic pixel-sensor prototypes featuring the AMS aH18 (180 nm) commercial CMOS technology. By fully integrating both analog and digital readout-circuitry on the same particle-detecting substrate, current challenges of hybrid sensor technologies, i.e., larger readout input-capacitance, lower production-yield, and higher production and integration cost, can be downscaled. The large electrode design using high-resistivity substrates actively helps to mitigate the charge-trapping effects, making these chips radiation hard. The surface and bulk damage induced in high irradiation environment change the effective doping concentration of the device, which modulates high electric fields as the reverse-bias voltage increases. This effect can cause high leakage current and premature electrical breakdown, driven by impact ionization. In order to assess the characteristics of heavily irradiated samples, we have carried out dedicated campaigns on ATLASPix1 chips that included irradiations of neutrons and protons, made at different facilities. Here, we report on the electrical characterization of the irradiated samples at different ambient conditions, also in comparison to their pre-irradiation properties. Results demonstrate that hadron irradiated devices can be safely operated at a voltage high enough to allow for high efficiency, up to the fluence of $2\times10^{15}$ $n_{eq}/cm^2$, beyond the radiation levels (TID and NIEL) expected in the outermost pixel layers of the new ATLAS tracker for HL-LHC.

[1] Corresponding author.

## Contents

## 1. Introduction

The upgrades of the ATLAS experiment at the High-Luminosity LHC (HL-LHC) aim at the complete replacement of their current tracking detectors to cope with the predicted higher event pile-up (200 events/bunch-crossing) and integrated luminosity of 4000 $fb^{-1}$. Detectors of Innermost Tracking (ITk) layers will need to withstand very large radiation fluences up to $1.3\times10^{16}$ 1-MeV equivalent neutrons per square centimeter ($n_{eq}.cm^{-2}$), and they are planned to be replaced within an intermediate shutdown of HL-LHC. However, the outer layer pixel matrices shall receive relatively lower cumulated fluence $1.5\times10^{15}$ $n_{eq}.cm^{-2}$ with a safety factor of 1.5 [1]. HV-CMOS sensors owe intrinsic properties of Monolithic Active Pixel Sensors (MAPS) featuring higher granularity (using standard CMOS processes), low-power consumption, lower system-level cost, increased data processing and high radiation tolerance [2], denoting them as superior candidates for the outer layer pixel matrix application. However, despite their remarkable performance in non-irradiated conditions, the extreme radiation tolerance demands of HL-LHC is calling for the development of a new generation of these devices which should feature a larger depletion depth using high-resistive substrates.

AMS AG of Austria is one of the fabrication facilities pioneered in the development of HV-CMOS sensors, has already shown a demonstrated depletion depth increase greater than 100 µm for 350 nm prototypes fabricated at high resistivity Si-substrate [3]. The successor prototype AMS aH18 ATLASPix, designed for 180 nm CMOS feature size, comes with a similar large deep N-Well electrode design integrating analog pixel electronics while maintaining two different novel readout schemes, the column drain readout scheme and triggered readout scheme. A limited number of samples from the first pre-production batch (aimed for process qualification) of ATLASPix1 arrived at the end of 2017. The final batch of the same process received later at the beginning of the 2$^{nd}$ quarter of 2018. The extensive electrical characterizations made at wafer and die-level helped to understand the intrinsic behavior of leakage current and breakdown voltage to a large extent.

In this paper, we report on the instrumentation procedure performed on the ATLASPix1 pixelated sensors, as well as on the results from the electrical characterization before and after irradiation with neutrons and protons. Preliminary results from the electrical characterization made on the successive generation ATLASPix2 at the nonirradiated case is also reported here showing the process qualification between AMS AG and TSI Semiconductor foundries.

## 2. Fabrication of HV-CMOS

The devices presented in this paper are the first ATLASPix HV-CMOS of 180nm feature size on Magnetic Czochralski (MCz) p-type Si-substrate ever fabricated. MCz wafer has chosen a great compromise to the foreseen radiation hardness of HL-LHC type accelerator, low leakage current, and cost. This type of wafer helps to maintain a more substantial contribution in mitigating the metallic impurity effect (through enriched oxygenated vacancies) and reduces slip defect during processing in different temperature-cycles.

However, MCz also holds few drawbacks over Float Zone (FZ) wafer processing like the macroscopically non-uniform radial distribution of dopant and impurities from the ingot, leading to lower resistivity for wafers greater than 200 - 300 mm diameter. Several new processes are developed for higher resistivity, but they are still commercially viable up to 1000 Ω.cm. Besides, oxygen complexes can form at MCz wafers between 400 °C – 600 °C that are known as Thermal Donors (TD). These thermal donors act as a great source of leakage current in a device [4].

## Process design

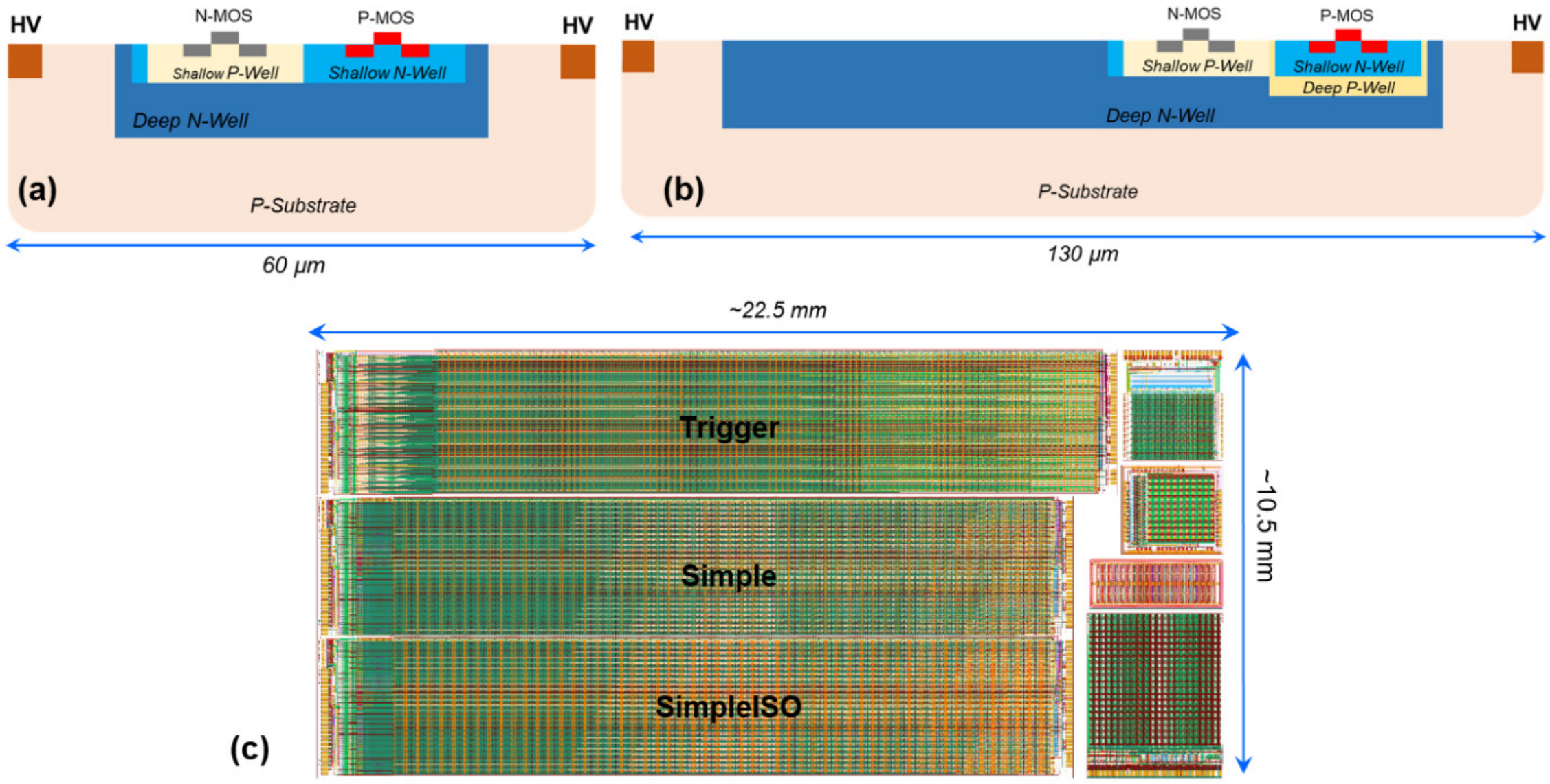

**Fig. 1**: Schematic cross-section of ATLASPix1: (a) Trigger matrix has dimension 60 µm × 50 µm, (b) Simple trigger-less matrix has dimension 130 µm × 40 µm. Deep P-Well isolating P-MOS is only present in SimpleISO matrix flavor. Dimension showed above are in relative scale in the longitudinal direction. (c) ATLASPix1 top layout reticle is holding all three-pixel flavors in the single diced die, used in RO system for wire bonding, share the same p-type substrate.

The schematic cross-sections of the proposed triggered and trigger-less readout of ATLASPix1 are shown in fig. 1 (a) & (b). The basic structure shown in the figure consists of a deep N-Well that is used to support two primary functions, acting as i) the charge collection electrode and ii) the substrate for PMOS transistors. The pixel analog electronics have been embedded inside the deep N-Well. High voltage node placed outside the deep N-Well is used to deplete the P-Substrate, an essential step of drifting charged particle produced from the ionizing radiation. Charge sensitive amplifier embedded in pixel analog electronics module amplifies the signal, which is later compared to a fixed discriminator threshold before sending to Read-Out (RO) electronics (situated at an edge of the sensor). Several flavors included during the design choice in terms of pixel geometry. The trigger-less option comes with a larger pixel area 130 µm × 40 µm, as shown in fig. 1(b), denotes as Simple matrix of 25×400 pixels. The N-MOS transistor present in Simple matrix has P-Well as a substrate, lies within the greater N-Well region, may tend to consume more power during operation and can lead devices to be less radiation hard. To address the issue, an additional flavor of Simple matrix implemented with CMOS-type comparator isolating shallow N-Well of P-MOS within a deep P-Well, known as SimpleISO. The third kind of matrix flavor has trigger RO scheme embedded, named as Trigger matrix. It holds the similar design of Simple matrix, has a smaller pitch of 60 µm × 50 µm (fig. 1(a)), consists of 56×320 pixels. A layout of ATLASPix1-reticle shown in fig. 1(c) represents all three-pixel flavors as to be found in a die, sharing the same depleting substrate.

AMS AG used 200 mm wide MCz wafer of thickness 725 µm with different nominal substrate resistivity flavors: 20, 80 and 200 Ω.cm. Each processed wafer holds 52 working reticles, each being of ~ 2 × 2.5 $cm^2$. Each reticle holds two pixelated HV-CMOS sensors: one is dedicated for Mu3e experiment [5] and another for ATLAS experiment. Top corner of the reticle holds several test chips, i.e., TCT test structure, CCPD chip for CLIC experiment, etc.

## 3. Experimental Setup and Irradiation Campaign

The experimental setup consists of an integral system of Cascade Microtech CM300 semi-automatic probe station and Advance Temperature Test (ATT) systems, situated in University of Geneva (UniGe). Measurements were made at different temperatures between +20 °C and -20 °C with 10 °C gradient while the dew point is maintained ~ -50 °C. A Keysight B1500A parameter analyzer was used as the voltage source that facilitates four simultaneous Source/Measure Units (SMU) probing at a range of ± 200 V with ±100 μV accuracy. Another Keysight B2200A logical matrix used in addition as the lower sensitive leakage current measurement unit has accuracy ±10.6 fA only. Thermal chuck absolute temperature of Cascade Microtech 300mm deviates ±2 °C during each investigation cycle. A large Programmable Logic Control (PLC) bit (>20 bits) was used for acquiring the leakage current below nA range, made each investigation cycle few 10 mins long. All these instrumental uncertainties have been carefully taken into account at the measured data.

Neutron irradiations were performed for ATLSPix1 candidates at the TRIGA Mark II reactor at the Jozef Stefan Institute – JSI (Ljubljana, Slovenia). TRIGA Mark II is a light-water reactor fueled by solid elements with a maximum power of 250 kW. Several tubes in the reactor core can be used for irradiation purposes, and the reactor power can be varied to achieve fluxes up to $4\times10^{12}$ $n.cm^{-2}.s^{-1}$. The neutron energy spectrum is broad, ranging from thermal neutrons up to 10 MeV. The hardness factor is 0.9, and the accuracy in the neutron fluence is given as 10% [6]. The fluence intensity of beam has gradient 4% per ~ 4 $cm^2$, which would have a minor effect at our ATLASPix1 prototypes.

Several other HV-CMOS sensors have also been irradiated with 16.7 MeV in BERN cyclotron to understand the damaging effect by charged hadrons. Coulomb interaction driven non-ionising energy loss (NIEL) by the charged particle creates the displacement of atoms and consequently affect in degrading the charge collection properties. The facility uses such low energy mainly for medical application with hardness factor 3.6. The beam profile is monitored using a dedicated collimator plate and crosschecked with 300 μm dosimetric film. Energy estimated as exposed in the Device under Test (DUT) has the uncertainty of 6%, and the beam intensity deviates 20% from the center on a 4 $cm^2$ exposed area [7].

## 4. Simulation and electrical characterization

In order to predict and optimize the performance of these ATLASPix HV-CMOS, numerical device simulations have been performed on Simple pixel pitch using Synopsis TCAD software. To add simplicity in simulation, ~6 μm deep N-Well (without any P-Well placed inside) along with the innermost metal structure for N-Well and HV contact (as in fig. 2 (a)) are considered preliminary, while values of all relevant parameters are representative of AMS 180 nm IBM process technology.

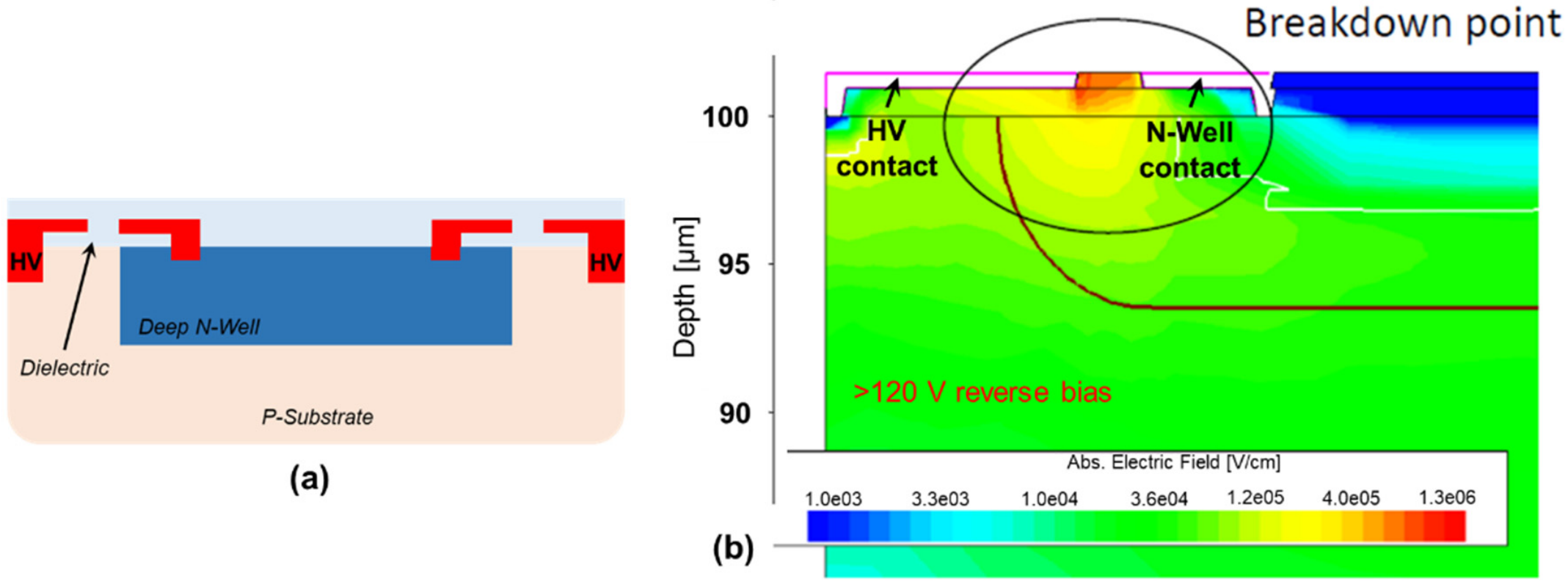

**Fig. 2**: The simulated (a) schematic cross-section for TCAD (scale in relative) and (b) result of p-n diode breakdown voltage.

Full 3D simulations were performed, also incorporating detailed information about the dielectric layers, showing the breakdown voltage as large (>120 V). Preliminary inspections of E-field distribution at a voltage close to breakdown yields similar peak values at the surface diode junction between HV and N-Well metal contact (fig. 2 (b)). A more well-defined simulation step including P-MOS and N-MOS by AMS process parameterization shall be explored in foreseen days. The relevant surface and bulk damage will also be explored afterward.

### 4.1 ATLASPix1 characterization

#### Before irradiation

Many ATLASPix1 dies were probed with the experimental setup in a precise ambient condition for electrical characterization at different chuck temperatures, using the required combination of Keysight 1500A parameter analyzer and Keysight B2200A logical matrix through GPIB interface. Semi-automatic Cascade CM300 needs the manual wafer loading only, where the entire probing steps can be made automatically with software scripts.

I-V curves of all three-pixel sensors flavors were also measured on the diced dies. Applied Dicing Before Grinding (DBG) process on wafers kept the ~50 µm cut line distance from the last edge-pixel pitch in an aim to keep a low dead periphery. An aggressive process choice with lower distance also made to slim-edge where only an AMS design kit generated singular guardring implemented. The samples thinned down to 50 – 100 µm thickness depending upon the substrate resistivity and do not hold post backside p-implant deposition and metallization. The depleted volume in HV-CMOS increases laterally near the $SiO_2$/Si interface and vertically toward the P-Substrate with the reverse bias increase. The dicing width and thickness should be carefully chosen; it can be a challenging issue for a highly resistive substrate wafer, where depleted region near the surface reaches the crystal defects at the cutting-edge and aggregate the leakage current at a massive scale [8]. Table 1 represents the greater details of the samples considered within the investigations. To understand the intrinsic behavior of the matrices and qualify the prototype design and technological process, I-V measurements were made at different ambient conditions.

**Table1**: Details of investigating ATLASPix1 Samples. Here, GR denotes guardring.

| Sensor Wafer ID | Substrate Resistivity (Ω.cm) | Nominal Resistivity (Ω.cm) | DBG Dicing Thickness (µm) | Chosen Cut-Edge to GR Distance (µm) |
|---|---|---|---|---|
| W06 | 10 - 20 | 20 | 100 ± 5 | 50 |
| W07 | 50 - 100 | 80 | 100 ± 5 | 50 |
| W09 | 50 - 100 | 80 | 70 ± 10 | 30 |
| W10 | 50 - 100 | 80 | 70 ± 10 | 30 |
| W22 | 100 - 400 | 200 | 100 ± 10 | 30 |
| W23 | 100 - 400 | 200 | 100 ± 10 | 30 |

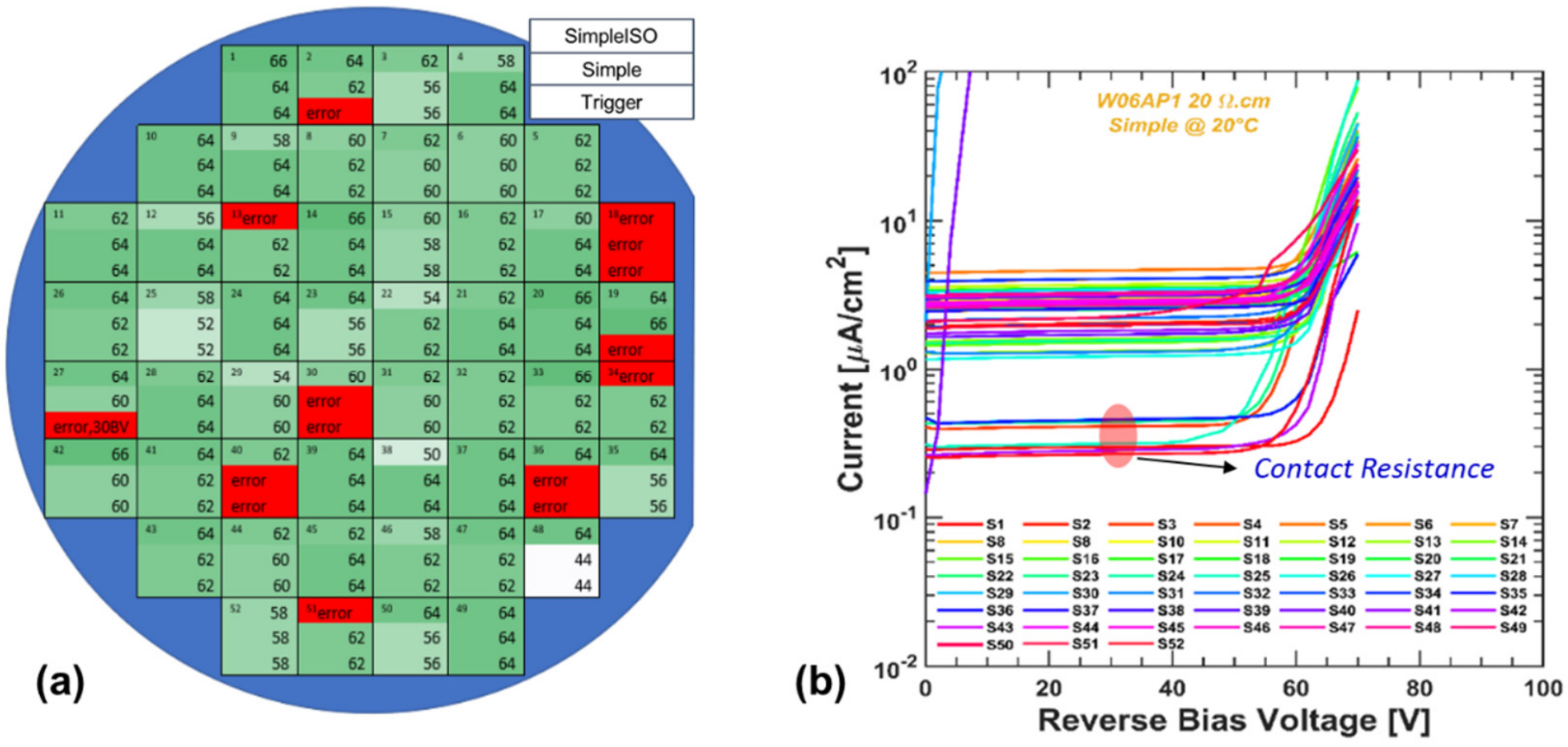

**Fig. 3**: (a) Measured electrical investigations on an ATLASPix1 W06 wafer, has nominal resistivity 20 Ω.cm and (b) I-V results normalized to the area of the simple matrix from 52 reticles of the wafer. Both measurements are performed in a precisely controlled ambient condition at 20°C.

As an example, fig. 3(a) shows the results of the electrical investigation made on all three matrices of the W06 wafer that has substrate resistivity 20 Ω.cm. The red error blocks in the same figure denote the pixel matrices to have an early breakdown (<20 V). In each reticle, the left inset text indicates the reticle number while the right texts are representing the reverse breakdown voltages of SimpleISO down to Trigger. During the I-V investigations, all power lines of RO electronics have kept with design recommended potentials: VDDD (1.8 V), VDDA (1.8 V) and VSSA (1 V). The breakdown voltage spreads between 50 - 62 V that is almost half from the simulated expectation. Placing probing pads outside of pixel matrix without the optimized guarding-design trenches them to act as parasitic MOS structure, accumulates process-quality driven surface generation current from the periphery [9] and so may trigger to an earlier avalanche breakdown at relatively low reverse bias applied on HV lines. Aside layout design improvement, the possible post-sintering process at the wafer level at 420 °C for a long time (i.e., ~80 - 120 mins) can reduce the TD quantity and lead to improvements in the leakage current.

I-V characteristic of all Simple matrices of 52 reticles is shown in fig. 3(b) where the maximum leakage current is 5 μA/cm$^2$ at 20 °C. The leakage current is expected to be one order of magnitude lower at -10 °C. However, if the measured data of 20 °C is normalized to the pixel number in a matrix, the per-pixel leakage appears as less than 5 nA, well below the HL-LHC design requirement (10 nA/pixel) [1]. Several matrices showed the early breakdown, driven by the processed induced point-defect. Some I-V data also showed one order of magnitude lower leakage current, which can be explained by wafer surface flatness. The optical inspection made later at the wafer surface in CM300 showed ~10 - 20 μm variance that drove into the probe-needle loose contact during automatic-measurement, resulted in additional contact resistance. Nevertheless, the yield of electrically functional chip per wafer is found to be greater than 70%.

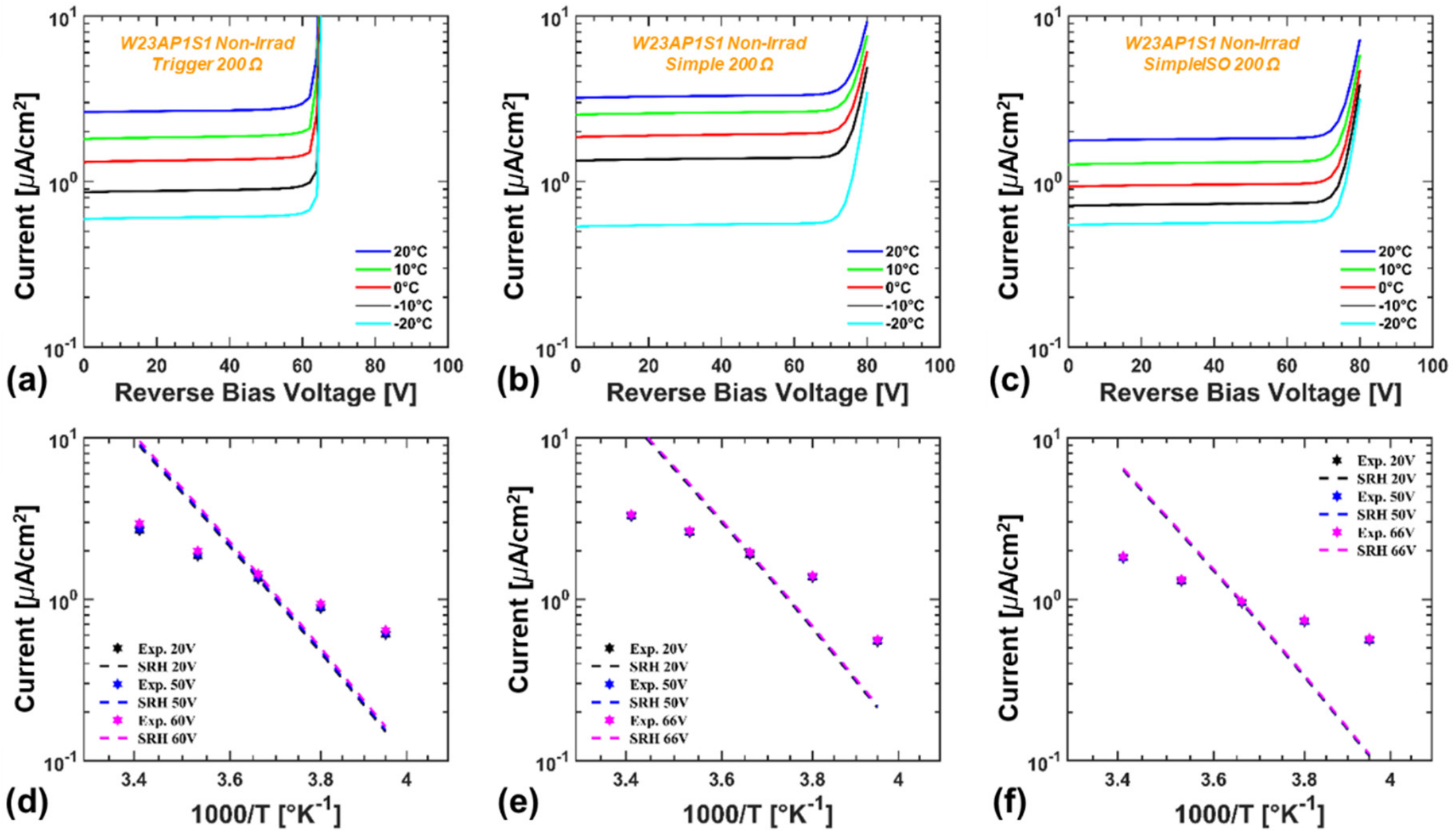

**Fig. 4**: Measured I-V curves at different ambient conditions of different pixel flavors: (a) Trigger, (b) Simple, and (c) SimpleISO of W23 wafer. The consequent Arrhenius plots of different pixel flavors: (d) Trigger, (e) Simple, and (f) SimpleISO.

As an example, fig. 4 represents I-V curves of all three-pixel flavors of ATLASPix1 thin prototypes of the nominal substrate resistivity 200 Ω.cm of W23 wafer. Given that the depletion width is proportional to reverse bias voltage and inversely proportional to the donor concentrations, we expect the high substrate resistivity complies to be more prone in collecting the peripheral current. The depletion region underneath the outermost guarding can be extended up to the saw-cut line of diced matrix-edge or grinded back surface, would explode the leakage current-scale where there are many crystal defects. Irrespective to pixel flavors, the leakage current remains ~1 μA/cm$^2$ and breakdown voltage is around 60 V, as reported in fig. 4 (a), (b) and (c). Arrhenius plot is an essential tool to understand the leakage current at different ambient temperatures, generated at the depleted bulk volume. The Fermi energy level in the material changes with temperature and helps to push electrons to conduction level; higher temperature also plays a role in enhancing the intrinsic dopant concentration through the thermionic emission process. There have been clear disagreements of the measured leakage data in all three bias references with the Shockley Read-Hall (SRH) calculation (is used to quantify the thermally generated carrier) as of fig. 4 (d), (e) and (f). In the calculation, data obtained at different ambient temperatures scaled to the values measured at 0 °C and also considered the effective band-gap energy 1.21 eV [10]. These disagreements clearly denote that the peripheral current vastly dominates on the measured leakage current.

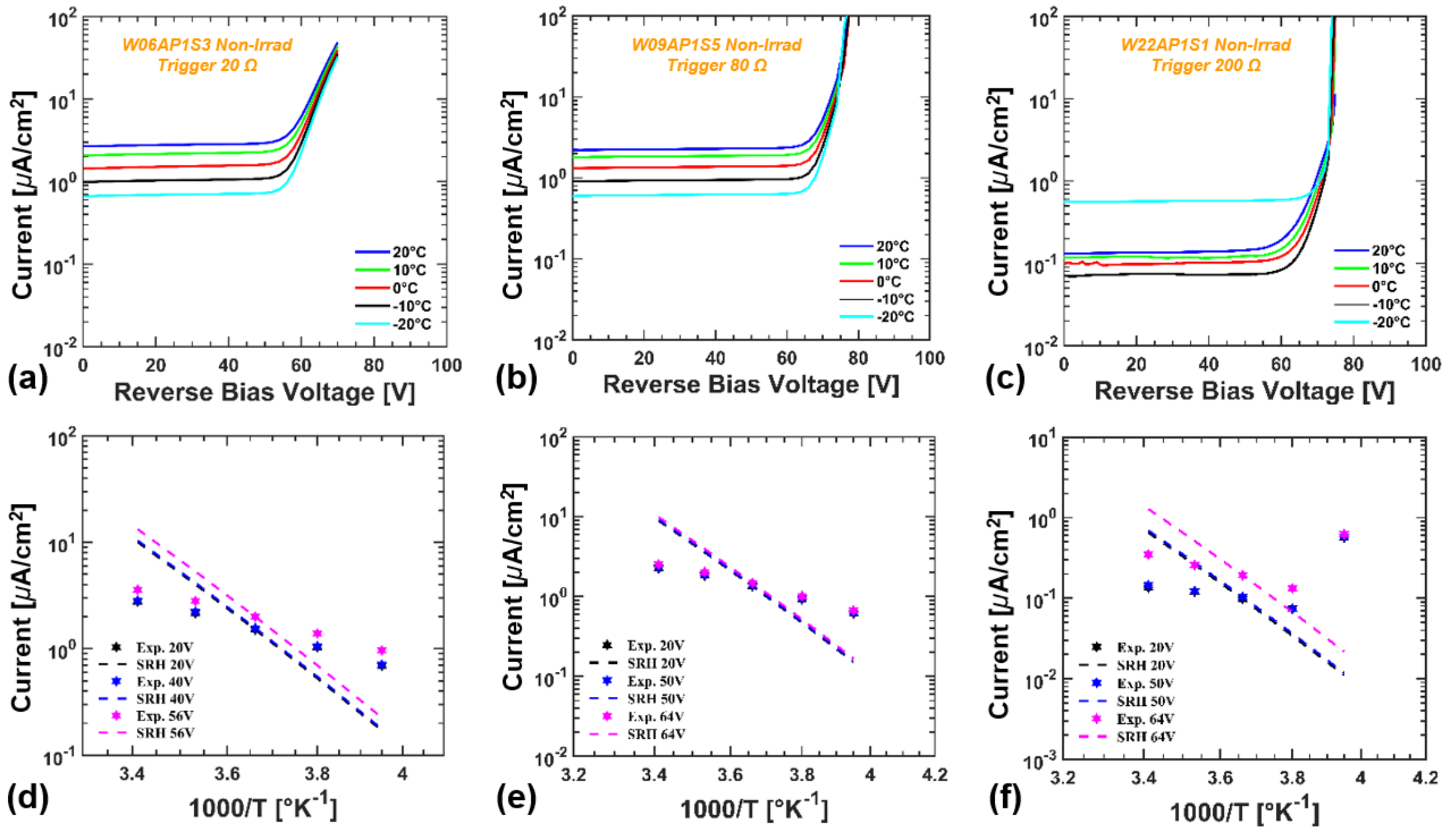

**Fig. 5**: Measured I-V curves of Trigger matrix at the different ambient conditions of different wafers: (a) 20 Ω.cm of W06, (b) 80 Ω.cm of W09, and (c) 200 Ω.cm of W22. (d), (e) and (f) present the Arrhenius plots of the Trigger matrix of the respective P-Substrate resistivity.

Fig. 5 (a), (b) and (c) show I-V curves for selected samples of diced Trigger matrix of three different resistivities: 20, 80 and 200 Ω.cm from W06, W09 and W22 wafers respectively. Since the low resistive p-type substrate holds higher donors, could show the leakage current increase with higher substrate resistivity at a particular reverse bias. However, a leakage current shows almost similar, ~1 μA/cm$^2$ at -10 °C for substrate flavors: 20 and 80 Ω.cm. For W22 wafer sample, the reported leakage current is even lower and thus indicates the surface condition is better in comparison to wafers W06 and W09. However, the Arrhenius disagreement for the bias reference near avalanche induced breakdown in all three cases still seconds the surface leakage domination. The breakdown voltage remains in these three prototypes ~60 V. Unfortunately, a careful observation from fig. 5(c) shows the abnormal leakage current rise at lower ambient condition (-20 °C); the poor dicing quality can explain it. Optical inspection made at die-level of the same wafer shows the dicing edge-cut distance from the loner guardring is very less than 30 μm or almost destroyed the guardring in most cases. At a relatively low temperature, the mean free path of charge carrier increases [11], and thus the diffusion-limited leakage can enhance. Massive leakage current may also aggregate from the dicing edge in a case where the lateral depletion region extends beyond the guard ring present in these samples, having substrate resistivity 100-400 Ω.cm. We believe, this abnormal effect would be negligible as denoted in [12] for overall sensor-operation; the reported leakage current rise is well below of 1 μA/cm$^2$ before irradiation. The situation can even improve after irradiation where the larger interface trap quantity takes hold of such leakage injection from the diced-edge. However, an important point suggests here that a cut-edge distance from the gaurdring at the dicing stage should be greater than 50 μm to avoid such anomaly. Please note, the sample showed a similar abnormal leakage behavior was not considered for any irradiation or the test-beam campaign later.

### After irradiation

Few selective and electrically qualified ATLASPix1 sensors of W23 wafer were irradiated at JSI neutron at three different fluences: $5\times10^{14}$, $1\times10^{15}$ and $2\times10^{15}$ $n_{eq}/cm^2$. To mitigate the annealing effect, irradiated samples were kept in freezing condition.

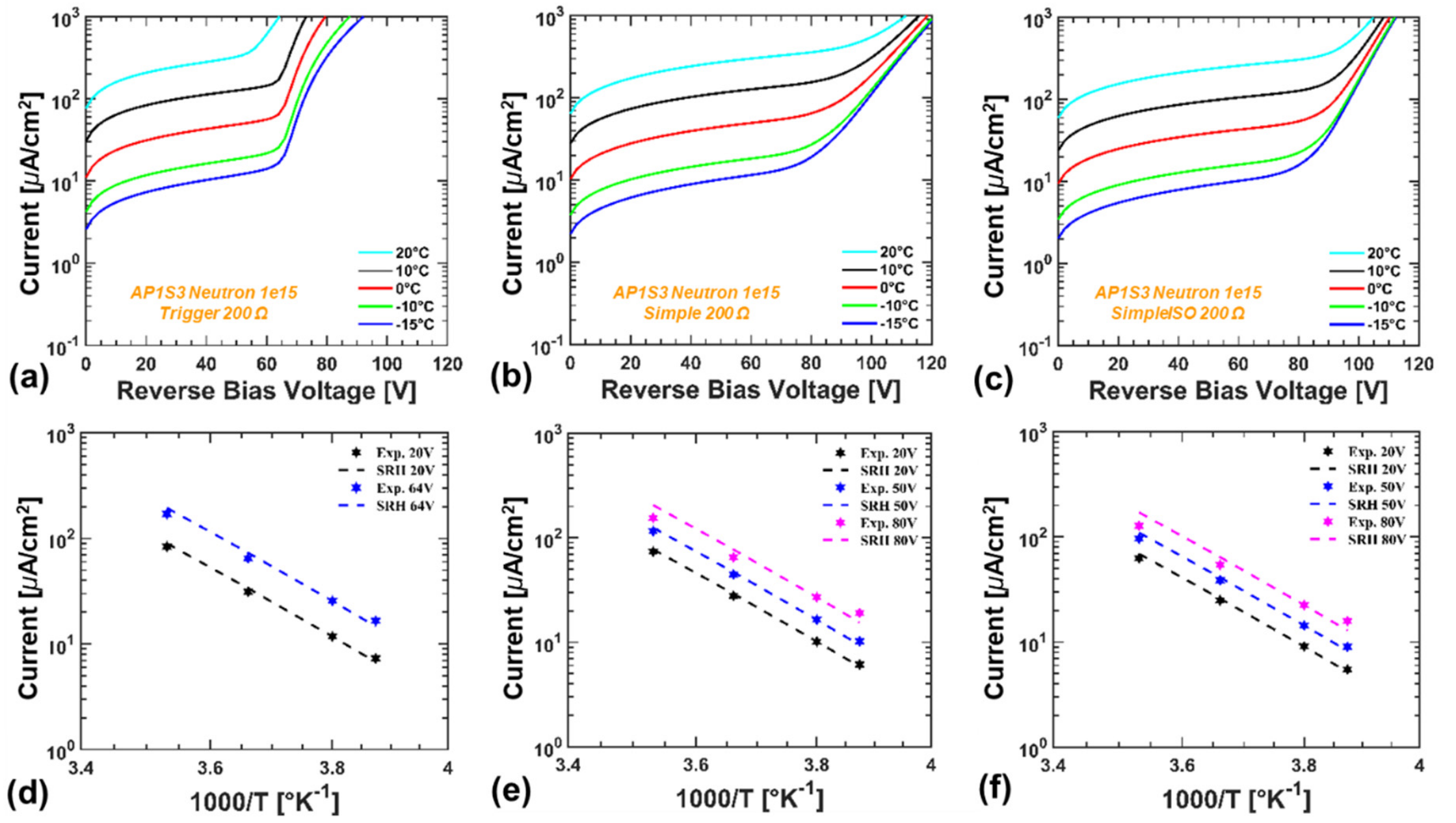

**Fig. 6**: I-V curves of different matrices of 200 Ω.cm substrate resistivity of W23 wafer at the different ambient conditions: (a) Trigger, (b) Simple, and (c) SimpleISO, irradiated with JSI neutrons at $1\times10^{15}$ $n_{eq}/cm^2$. (d), (e) and (f) present the Arrhenius plots of the respective pixel flavors.

**Table 2**: Summary of the electrical characteristics of ATLASPix1 samples of W23 wafer irradiated with neutrons at JSI.

| Fluence [$n_{eq}/cm^2$] | TID [Mrad(Si)] | Device ID | $J_{lk}$ at -10°C [μA/cm²] | α* [$10^{-17}$ A/cm] | $V_{bd}$ [V] at -10°C |
|---|---|---|---|---|---|
| $5\times10^{14}$ | 0.56 | JN-AP1S01-5e14-200 Trigger | 11.62±2.24 @68V | 3.13±0.42 @56V | 70±2 |
| | | JN-AP1S01-5e14-200 Simple | 13.11±2.31 @70V | 3.51±0.76 @62V | 72±2 |
| $5\times10^{14}$ | 0.56 | JN-AP1S02-5e14-200 Trigger | 10.98±2.24 @68V | 3.07±0.41 @56V | 70±2 |
| | | JN-AP1S02-5e14-200 Simple | 11.99±2.31 @76V | 3.56±0.60 @70V | 78±2 |
| | | JN-AP1S02-5e14-200 SimpleISO | 12.29±2.31 @72V | 3.55±0.59 @66V | 74±3 |
| $1\times10^{15}$ | 1.12 | JN-AP1S03-1e15-200 Trigger | 25.53±2.24 @64V | 3.30±0.45 @50V | 66±2 |
| | | JN-AP1S03-1e15-200 Simple | 29.68±2.32 @82V | 4.16±0.83 @76V | 84±2 |
| | | JN-AP1S03-1e15-200 SimpleISO | 24.43±2.32 @82V | 3.52±0.71 @76V | 84±2 |
| $1\times10^{15}$ | 1.12 | JN-AP1S04-1e15-200 Trigger | 24.33±2.24 @66V | 2.99±0.37 @50V | 68±2 |
| | | JN-AP1S04-1e15-200 Simple | 25.25±2.32 @82V | 3.66±0.69 @76V | 84±2 |
| | | JN-AP1S04-1e15-200 SimpleISO | 25.16±2.32 @82V | 3.62±0.74 @76V | 84±2 |
| $2\times10^{15}$ | 2.24 | JN-AP1S05-2e15-200 Trigger | 34.69±2.24 @62V | 2.27±0.32 @50V | 64±2 |
| | | JN-AP1S05-2e15-200 Simple | 39.72±2.32 @88V | 2.64±0.38 @80V | 90±2 |
| | | JN-AP1S05-2e15-200 SimpleISO | 41.23±2.32 @90V | 2.65±0.38 @80V | 92±2 |

As an example, fig. 6 shows the I-V curves and the respective Arrhenius plots of ATLASPix1 pixel flavors, irradiated up to $1\times10^{15}$ $n_{eq}/cm^2$ and measurements made at different ambient temperatures. In all pixel flavors, the leakage current increased one order of magnitude higher than data reported already for the non-irradiated sample. While the breakdown voltage of Trigger matrix remained similar as before irradiation, the breakdown voltages for Simple and SimpleISO increased beyond 80 V reverse bias, which can be anticipated from the pixel geometry difference between trigger and trigger-less pixel architectures. Trigger-less pixels consist of a wider interface-state region between N-Well electronics to HV line (as in fig. 1(b) at longitudinal direction) where the E-field distribution at a respective reverse bias is relatively weaker. Neutron irradiation induced bulk and interface traps mainly seize the charge carriers which requires additional potential applied in these pixel matrices before they go to avalanche breakdown.

Table 2 summarizes the most significant data for the ATLASPix1 neutron-irradiated samples. Minor differences at the breakdown voltage of trigger pixel architecture are observed among different irradiated fluences. The values of the current density ($J_{lk}$) refer to a bias voltage close to the breakdown voltage, in order for the depletion volume to be as large as possible. This fact, along with the lack of dedicated edge-TCT measurement on these prototypes and other non-idealities (i.e., the uncertainties in the irradiation fluences, the self-heating effects experienced by the devices during irradiation) make the extraction of the current damage constant α less significant. The leakage current density of sample alone would not give an immediate idea of the order of magnitude, so we report the geometric current related damage rate (α*), as defined in [13] for dealing with sensors that cannot be fully depleted. The computed α*, having considered all-important uncertainties, shows a good agreement with the data reported in [14]. Hereby, values could differ from the standard α which has information on the actual depleted volume. The decrease of α* value at $2\times10^{15}$ $n_{eq}/cm^2$ can be ascribed from the surface damage effects by γ-ray increase from reactor background. The surface damage effects help to aggregate the peripheral current, and so trigger impact ionization process at a lower reverse bias. On the contrary, the breakdown increase with fluence at trigger-less matrices points the geometry-dependent E-field distribution where additional bias required to release trapped carriers.

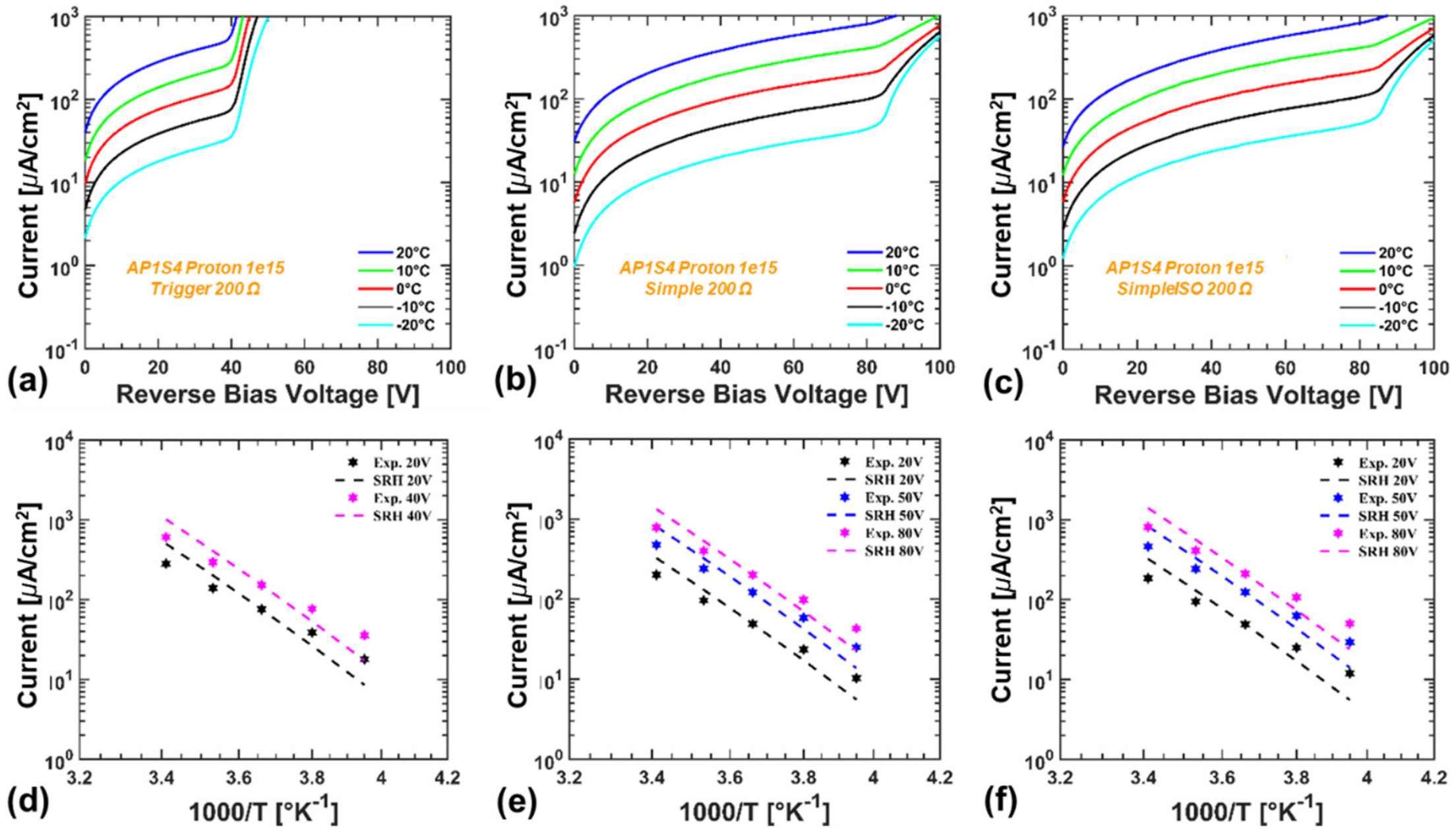

**Fig. 7**: I-V curves of different matrices of 200 Ω.cm substrate resistivity of W23 wafer at the different ambient conditions: (a) Trigger, (b) Simple, and (c) SimpleISO, irradiated with Bern cyclotron protons at $1\times10^{15}$ neq/cm2. (d), (e) and (f) present the Arrhenius plots of the respective pixel flavors at different bias references.

Several 200 Ω.cm ATLASPix1 candidates of W23 wafer were also irradiated at Bern cyclotron with 16.7 MeV protons. Similar fluence choice was made: $5\times10^{14}$, $1\times10^{15}$ and $2\times10^{15}$ $n_{eq}/cm^2$ to have a fair comparison with neutron irradiation. As expected, the leakage current increased in all pixel flavors after proton irradiation due to both surface and bulk damaging effects; the scale went up to almost two orders of magnitude higher in comparison to the non-irradiated case (fig. 7(c)). Interestingly, the breakdown voltage of trigger-less architecture decreased relatively to neutron irradiation at the highest fluence received, likely due to the much larger TID associated with proton irradiation [7]. The Arrhenius behavior of the pixel flavors reported in fig. 7(d), 7(e), and (f) at different bias references is showing better agreement to SRH calculation with respect to the non-irradiated case. This hints the damaging effect driven intrinsic leakage current rise was high enough to hinder the peripheral current, an already good indication for not being prone after irradiation to RO noise rise and thermal-runway. However, SRH agreement to measured data is not definitive as it was for neutron, indicating that larger surface damaging effect still acts a role to cumulate the peripheral current. Careful observation on fig. 7(a) shows the breakdown voltage of Trigger matrix decrease with temperature rise, denotes the peripheral current injection enhancement by the punch-through process and related consequences.

Table 3 summarizes the most important data for all samples of ATLASPix1 irradiated with protons. Considerations similar to those made for tables 2 apply to the α* values of table 3. Compared to the breakdown values of Trigger matrix reported at the table decreased with fluence. As it can be ascribed, the larger surface damaging effect trenched larger peripheral current and triggered the earlier avalanche breakdown.

**Table 3**: Summary of the electrical characteristics of ATLASPix1 samples of W23 wafer irradiated with protons at Bern cyclotron.

| Fluence [$n_{eq}/cm^2$] | TID [Mrad(Si)] | Device ID | $J_{lk}$ at -10°C [μA/cm$^2$] | α* [$10^{-17}$ A/cm ] | $V_{bd}$ [V] at -10°C |
|---|---|---|---|---|---|
| 5×10$^{14}$ | 53.34 | BCP-AP1S01-5e14-200 Simple | 59.66±2.33 @83V | 16.97±6.96 @75V | 84±1 |
| | | BCP-AP1S01-5e14-200 SimpleISO | 61.55±2.33 @83V | 17.08±9.10 @75V | 84±1 |
| 5×10$^{14}$ | 53.34 | BCP-AP1S02-5e14-200 Trigger | 70.22±2.25 @45V | 18.53±10.05 @40V | 47±1 |
| | | BCP-AP1S02-5e14-200 Simple | 65.00±2.33 @83V | 17.14±6.96 @76V | 84±1 |
| | | BCP-AP1S02-5e14-200 SimpleISO | 58.45±2.33 @82V | 16.95±8.03 @76V | 83±1 |
| 1×10$^{15}$ | 106.69 | BCP-AP1S03-1e15-200 Trigger | 51.39±2.25 @40V | 7.95±3.90 @36V | 42±2 |
| | | BCP-AP1S03-1e15-200 Simple | 54.02±2.32 @82V | 8.33±4.32 @76V | 84±2 |
| | | BCP-AP1S03-1e15-200 SimpleISO | 119.72±2.34 @84V | 11.72±6.06 @76V | 85±1 |
| 1×10$^{15}$ | 106.69 | BCP-AP1S04-1e15-200 Trigger | 54.92±2.25 @40V | 8.47±4.58 @35V | 41±1 |
| | | BCP-AP1S04-1e15-200 Simple | 84.93±2.33 @84V | 12.15±5.88 @80V | 85±1 |
| | | BCP-AP1S04-1e15-200 SimpleISO | 92.16±2.34 @84V | 13.02±7.16 @80V | 85±1 |
| 2×10$^{15}$ | 213.37 | BCP-AP1S05-2e15-200 Trigger | 54.92±2.25 @35V | 2.94±1.24 @30V | 36±1 |
| | | BCP-AP1S05-2e15-200 Simple | 83.29±2.33 @86V | 5.47±2.58 @80V | 87±1 |
| | | BCP-AP1S05-2e15-200 SimpleISO | 82.34±2.33 @83V | 5.78±3.01 @77V | 84±1 |

### 4.2 ATLASPix2 characterization

ATLASPix2 HV-CMOS prototyped with a similar design of ATLASPix1 Trigger RO architecture, it has a 24×36 pixel matrix and has been fabricated at AMS AG with 180 nm standard CMOS IBM process (P-Substrate resistivity, 20 Ω.cm). Each pixel dimension has changed to 128×50 μm$^2$. This prototype aims to qualify various circuit blocks intended for the chip periphery. The Main Pixel Matrix (MPM), as highlighted with yellow color in fig. 8(a), is mainly designed to provide incentives for the trigger buffers located at the end of columns. Pixel Memory Array (PMA) as seen in red rectangular inset contains similar SEU-tolerant memory of MPM, dedicatedly designed for Single Event Effect (SEE) tests. Another ATLASPix2 submission was made to TSI Semiconductor, USA (the similar design of AMS ATLASPix2) to study the process quality at this other foundry. Both batch of ATLASPix2 productions completed in the last quarter of 2018.

After wafers reception, the following thinning and dicing steps were carried out. AMS production thinned down to 220 μm while the thickness of TSI production was kept at 254 μm. Individual diced dies were then carefully electrically characterized in UniGe facility with the same instrumental setup used for ATLASPix1 candidates. Thereby, the reported I-V data presented in fig. 8(c) for AMS ATLASPix2 and fig. 8(d) for TSI ATLASPix2 include the same experimental uncertainties. It is worth to stress here that there has been an issue involved on dry air refurbishment in the clean room lately and so, the ambient temperature varied between 30°C to 10°C with 5°C gradient to avoid the dew factors.

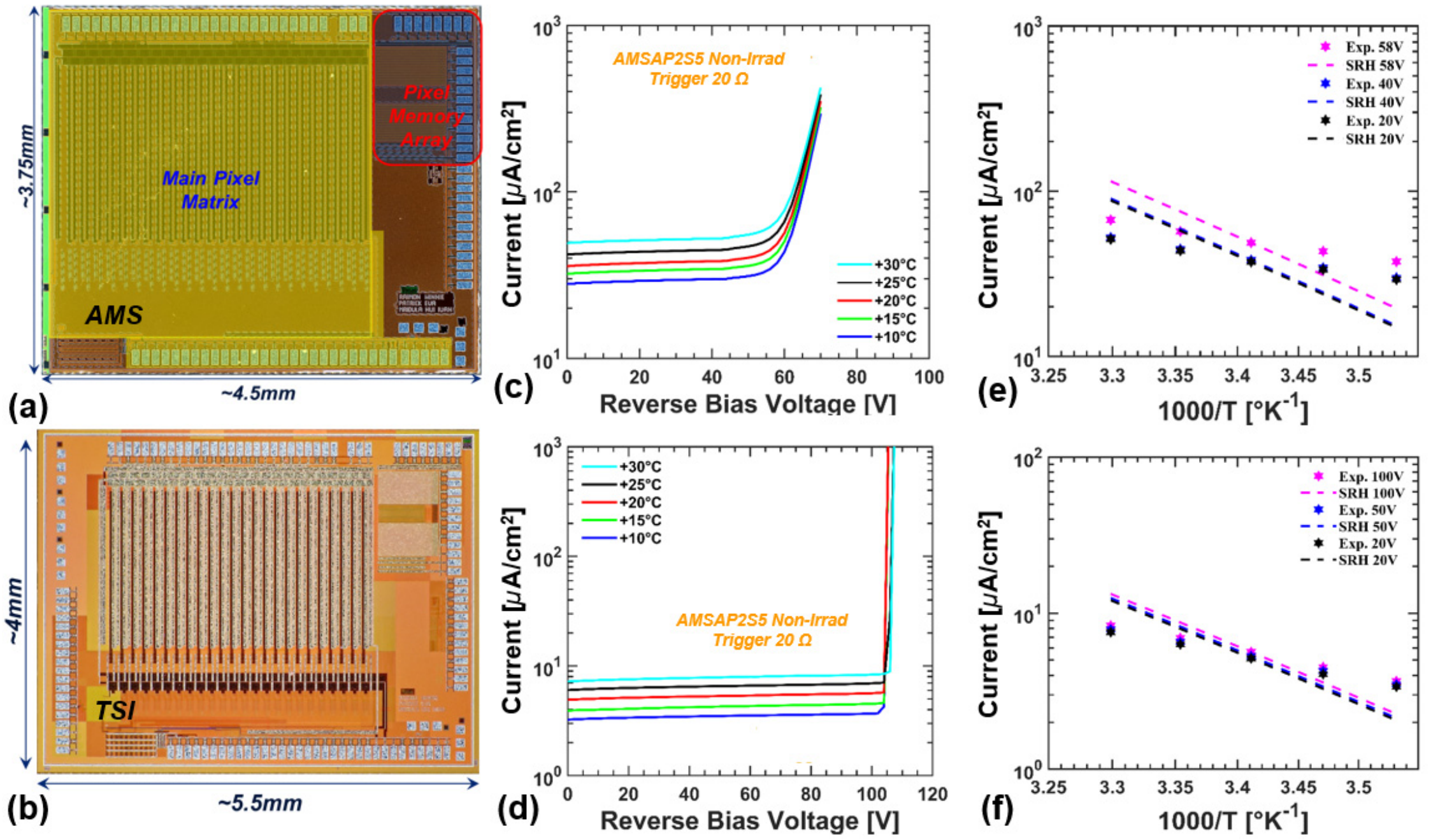

**Fig. 8**: (a) and (b) hold the respective micrograph of ATLASPix2 prototypes fabricated at AMS AG and TSI semiconductor. I-V curves show non-irradiated results of (c) AMS-ATLASPix2 and (d) TSI-ATLASPix2 at different ambient conditions. (e) and (f) present the respective Arrhenius plots of ATLASPix2 prototypes of AMS and TSI.

Like ATLASPix1, I-V data also have taken by enabling the electronics power lines: VDDD (1.8 V), VDDA (1.8 V), and VSSA (1.0 V). The leakage current in the junction between N-Well and P-Substrate is generally well maintained in most CMOS foundries nowadays through the matured doping profiles. However, Short Channel Effect (SCE) driven leakage at CMOS circuitry, that is almost two or three orders of magnitude higher than P-N junction leakage, depends vastly on foundry limitations: lithography resolution, ultra-fine oxide monolayer optimization, super steep wells, and halo implant growth techniques. I-V data reported in fig. 8(c) for AMS-ATLASPix2 prototype show the typical breakdown beyond 50 V reverse bias as it was seen on its predecessors, ALTASPix1. The measured leakage current of AMS-ATLASPix2 is a few 10 µA/cm$^2$ that would be at least one order of magnitude lower at -10 °C, still is in good agreement to the data measured in ATLASPix1. Fig. 8(d) reports the I-V plots for the same ATLASPix2 fabricated in TSI, showing eight times lower leakage current than samples fabricated in AMS, that hints to the better processing maturity of TSI Semiconductor and the lower surface damage at devices. The breakdown improvement beyond 100 V at the TSI process reclaimed expected values from TCAD simulation. Arrhenius plots of ATLASPix2 of different foundries: AMS (fig. 8 (e)) and TSI (fig. 8(b)) made at several reverse-bias references showed still the disagreement of SRH calculation (with reference to the measured value at 10 °C), may point to a significant contribution of peripheral current but we should also be careful since SRH prediction is over-estimated at a high-temperature reference [15]. Several layout improvements like optimized guard-ring adaption to the pixel-matrix boundary and inclusion of planar MOS and gated diode like test structures are to be accounted in the following submissions and so as the low-temperature investigations shall be explored.

## 5. Conclusion

We have reported on the electrical characterization of non-irradiated ATLASPix1 HV-CMOS processed in AMS AG. This study has included thorough investigations of prototypes from wafer to die differing in the trigger and trigger-less architectures. ATLASPix1 prototypes showed 1 µA/cm$^2$ leakage current at -10 °C, in compliance with the technical design requirement for the pixel sensors of HL-LHC. Several irradiation campaigns have made with neutrons at JSI and protons at Bern. The report focus has been mainly on high-field effects on the leakage current and breakdown voltage behavior, to specify the ATLASPix HV-CMOS optimal operating condition. Despite a non-uniformity found at measured data, the leakage current reacted to the theoretically expected increase with radiation fluence. The geometric current related damage rate $\alpha^*$ is comparable to the range of values typically observed in non-annealed samples within a large uncertainty due to the lack of exact depletion volume, variation of irradiation fluence and annealing conditions. Current measurements after irradiation at different temperatures are in good agreement with the SRH-based theoretical model, so that leakage current is confirmed most likely to be dominated by thermal generation in the depleted bulk up to the breakdown point.

Breakdown voltage exhibits interesting behaviors with fluence types, because of its more complex dependence on sensor geometry and fabrication quality. The breakdown voltage is found to increase after neutron irradiation in all samples, but not to the same extent. TID involved with proton irradiation seems like to decrease the breakdown of ATLASPix1 trigger flavor with fluence. The critical contribution of peripheral current to the Trigger matrix breakdown can be significantly improved if an effective E-field shield design and the relevant geometry optimization are imposed. Similarly to what was observed before irradiation, the breakdown voltage for trigger-less pixel flavors, Simple and SimpleISO, shows to be better even at relatively high TID values (~100 Mrad). Factually, these devices have a critical point for breakdown at the front between the N-Well and HV junction, where the distance between HV field plate and N-Well contact plays the role to modulate the electric field peaks and breakdown. The breakdown voltage values of ALTASPix1 seems to reach at ~90 V and remain functional for neutron fluence $2\times10^{15}$ $n_{eq}/cm^2$, beyond the accumulated fluence as it may experience in its lifetime at HL-LHC.

Preliminary electrical investigations reported here for the TSI process show already many encouraging results in comparison to AMS, especially the large breakdown (>100 V) and lower leakage. Foreseen considerations in design improvement and continuous instrumental qualification at current prototypes in coming days should make HV-CMOS as a competent choice for the outer pixel layers of HL-LHC.

## Acknowledgment

The authors would like to thank Prof. Vladimir Cindro and Dr. Igor Mandic of JSI, Ljubljana for their warm support to neutron irradiation. This project has received funding from the European Union's Horizon 2020 Research and Innovation programme under Grant Agreement no. 654168 and 675587. The research work presented in this paper was also supported by the SNSF grants 200021_169015, 200020_156083, 20FL20_173601 and 200020_169000.